\documentclass{ws-procs9x6}

\usepackage{amssymb}
 \def \C {\mathbb C }
 \def \R {\mathbb R}

 \def \U {1\! \! 1}

\def \be{\begin{equation}}
\def \ee{\end{equation}}
\def \bea {\begin{eqnarray}}
\def \eea {\end{eqnarray}}
\def \ba {\begin{array}}
\def \ea {\end{array}}
\def \bitz {\begin{itemize}}
\def \eitz {\end{itemize}}

\def \ben {\begin{enumerate}}
\def \een {\end{enumerate}}
\def \lb {\label}

\def \dis {\displaystyle}

\begin{document}

\title{Self-Adjoint Extensions of the
Dirac Hamiltonian with a $\delta$-Sphere Interaction}

\author{Gabriel Y.H. Avossevou$^{1,2}$,
Jan Govaerts$^{3}$\\
and Mahouton N. Hounkonnou$^{1,2}$}

\address{
$^{1}$International Chair in Mathematical Physics and Applications (ICMPA)\\
University of Abomey-Calavi, 072 B.P. 50 Cotonou, Republic of Benin\\
E-mail: avossevou@yahoo.fr, norbert\_hounkonnou@cipma.net}

\address{
$^{2}$Department of Physics, Faculty of Sciences and Technology (FAST)\\
University of Abomey-Calavi, Republic of Benin}

\address{
$^{3}$Institute of Nuclear Physics, Catholic University of Louvain\\
2, Chemin du Cyclotron, B-1348 Louvain-la-Neuve, Belgium\\
E-mail: govaerts@fynu.ucl.ac.be}

\maketitle

\abstracts{
The purpose of this paper is to make an explicit
construction of specific self-adjoint extensions of the Dirac
Hamiltonian in the presence of a $\delta$-sphere interaction of
finite radius. The exact resolvent kernel of the free
Dirac operator is given. This specifies related
results that have recently appeared in the literature.
}

\section{Introduction}

Solvable Hamiltonians with singular $\delta$-sphere potentials
have been the subject of many investigations in the past since
the work by Green and Moszkowsky.\cite{gm}
These models belong to the small class of models that are 
analytically solvable. As far as we know, all these studies have used
the von Neumann formalism of self-adjoint extensions
of symmetric linear operators. The earlier works concern
nonrelativistic quantum mechanics based on the Schr\"odinger
Hamiltonian with $\delta$ interactions. A good summary
of these nonrelativistic models is to be found in 
Refs.~\refcite{ags} and \refcite{hhs}. Comparatively, a much smaller
number of papers have dealt with the relativistic
$\delta$-Dirac Hamiltonian. To the best of our knowledge,
the first rigorous treatments of singular Dirac Hamiltonians
were provided by Dittrich, Exner and {\u S}eba.\cite{desm,des}
Next, in the same vein, there have been the papers by Avossevou 
and Hounkonnou,\cite{ah1,ah2,ah3,ah4,ah5} and more recently 
those by Shabani and Vyabandi\cite{sv} which include
quite an uncommon critique of the work in Refs.~\refcite{ah1}--\refcite{ah5}.
In this contribution, we provide self-adjoint extensions of the
$\delta$-Dirac Hamiltonian with support on a sphere of finite radius,
given specific boundary conditions.

The paper is organized as follows. We present the model
and give the normalised solutions for the free Dirac radial
operator in Secs.~2 and 3. This leads us to the
construction of the resolvent of the free Dirac radial
Hamiltonian in Sec.~4. In Secs.~5 and 6 we then
construct the resolvent of the $\delta$-Dirac Hamiltonian given
boundary conditions of the first type.

\section{Model and Boundary Conditions}

Let us consider in the Hilbert space ${\mathcal H}$
the formal expression describing the interaction model
\begin{equation}\label{M}
H = H_D + V(|\vec{x}|)\ \ ,\ \ 
\vec{x}\in \R^3\ ,
\ee
where $V(|\vec{x}|)$ is the interaction potential to be
specified hereafter.
%where ${\widetilde G}$ is {\it a priori\/} a $4\times 4$ matrix, to be 
%specified (see Eq.~(\ref{pot})). 
The quantity $H_D$ is the free Dirac operator,
\begin{equation}\label{fond}
H_D \equiv -i\hbar c{\bf {\underline{\alpha}}}\nabla +
{\bf {\underline{\beta}}}m c^2\ \ ,\ \ \hbar =h/2\pi\ ,
\end{equation}
where the following notations are used,
\begin{equation}\label{Dmat}
{\bf {\underline{\alpha}}}=\left(
\begin{array}{cc}
{\bf 0} & {\bf \sigma}\cr
{\bf \sigma} & {\bf 0}
\end{array}\right)\ \ \ ,\ \ \ 
{\bf {\underline{\beta}}}=\left( 
\begin{array}{cc}
\U_2 & {\bf 0}\cr
{\bf 0} & -\U_2 
\end{array}\right)\ ,
\end{equation}
${\bf \sigma}\equiv {\sigma}^\iota$ ($\iota= 1, 2, 3$)
being the $2\times 2$ Pauli matrices defined by
\begin{equation}\label{Pmat}
{\sigma}^ 1=\left(
\begin{array}{cc}
 0 & 1\cr
1 & 0
\end{array}\right)\ \ ,\ \ 
{\sigma}^ 2= \left(
\begin{array}{cc}
0 & -i\cr
i & 0 
\end{array}\right)\ \ ,\ \ 
{\sigma}^ 3= \left(
\begin{array}{cc}
1 & 0\cr
0 & -1 
\end{array}\right)\ ,
\end{equation}
so that the Hamiltonian is explicitly given by
\begin{equation}\label{matfond}
H_D=\left(
\begin{array}{cc}
mc^2{\U}_2 & -i\hbar c\sigma\cdot\nabla\cr
-i\hbar c\sigma\cdot\nabla & -mc^2{\U}_2 
\end{array}\right)\ \ ,\ \ 
{\U}_2 =
\left(
\ba{cc} 1 & 0\cr 0 & 1\ea
\right)\ .
\end{equation}
%The interaction potential associated to this decomposition
%writes as,\cite{desm}
%\be\lb{pot}
%V(|\vec{x}|)=G\delta (\vert \vec{x}\vert -R)=G\delta (r-R)\ \ ,\ \ 
%G=\left(\ba{cc}
%\alpha_o & 0\cr
%0 & \beta_o
%\ea\right)\ ,\ \alpha_o,\beta_o \in \R\ .
%\ee

The method adopted for the study of this Hamiltonian is
based on the von Neuman theory of self-adjoint extensions
of linear symmetric operators.\cite{aghkh,ag} The first step
is to consider within a Hilbert space ${\mathcal{H}}$ the symmetric operator
$\dot{H}\equiv H_D$,
${\mathcal{D}}(\dot{H})=\{\psi\in H^{1,2}(\R^3)\otimes \C^4 \;,
\; \psi(S_{R})=0 \}$,
where $S_R$ is the closed ball of radius $R$ centered at the
origin in $\R^3$, and $H^{1,2}$ the Sobolev space of indices $(1,2)$.
The closure of this operator is given by
\be
\overline{\dot{H}}=
\overline{H_D|_{C_0^\infty \left( \R^3\setminus
S_R\right)}}=
H_D|_{H_0^{1,2}\left( \R^3\setminus
S_R\right)}\ .
\ee
The Hilbert space ${\mathcal{H}}$ decomposes as\cite{desm,des}
\begin{equation}\label{DEC}
{\mathcal H}=\bigoplus_{j=(1/2)}^{\infty}\;
\bigoplus_{\ell=j-(1/2)}^{j+(1/2)}
 {U}_{j\ell}^{-1}\; \left[L^2((0,\infty), dr) \otimes \C^2\right]\bigotimes
\left[\Omega_{j\ell-j},\cdots ,
\Omega_{j\ell j}\right]\ ,
\end{equation}
where the isomorphism $U_{jl}$ has been introduced to
remove the weight factor $r^2$ from the measure
on the original radial space. The corresponding radial
self-adjoint operator writes as
\bea\lb{dcpn}\nonumber
{\dot H} = \bigoplus_{j=(1/2)}^{\infty}\;
\bigoplus_{\ell=j-(1/2)}^{j+(1/2)}
U_{j\ell}^{-1}{\dot h}_{j\ell}U_{j\ell}
\otimes{\U }\ ,
\eea
\be
{\dot h}_{j\ell} = \tau =
\left(\begin{array}{lc} mc^2    &\hbar c\left(-{\frac{d}{dr}} +
 \frac{\kappa_{j\ell}}{r}\right)\\ \hbar c\left(\frac{d}{dr} +
 \frac{\kappa_{j\ell}}{r}\right)     &-mc^2
\end{array}\right)\ ,
\ee
\bea
{\mathcal D}({\dot h}_{j\ell})=&& \left\{
\psi\left(\ba{l}f(r)\\g(r)\ea\right)\in L^2((0,\infty),dr)
\otimes\C^2;\;\psi\in AC_{loc}((0,\infty)\setminus \{R\});
\right.\nonumber\\
 &&\hspace{0.5cm}\left.{\dot h}_{j\ell} \psi\in L^2((0,\infty))\otimes
\C^2\right\}\ ,
\eea
where $f(r)$ and $g(r)$ determine the radial dependence
of the upper and lower Dirac bi-spinors. The symbol ${\U }$ in
Eq.~(\ref{dcpn}) stands for the unit operator.

Note that the radial operator obtained in Eq.~(\ref{dcpn})
is the same as the one given in Eq.~(2.9a) of Ref.~\refcite{desm}
provided a choice of units such that $\hbar=1=c$ is made.

The interaction potential associated to this bi-spinor decomposition
writes as,\cite{desm}
\be\lb{pot}
V(|\vec{x}|)=G\delta (\vert \vec{x}\vert -R)=G\delta (r-R)\ \ ,\ \ 
G=\left(\ba{cc}
\alpha_o & 0\cr
0 & \beta_o
\ea\right)\ ,\ \alpha_o,\beta_o \in \R\ .
\ee

The boundary conditions we use to characterize the 
self-adjoint extensions are obtained as follows.
Assume that the limits $\psi(R_\pm)$ exist and consider the
eigenvalue equation
\be
\left[{\dot h}_{j\ell}+ G\delta(r-R)\right]\psi =E\psi\ .
\ee
Then, integrating over $(R-\varepsilon, R+\varepsilon)$
taking the limit $\varepsilon\rightarrow 0$, and using
the relation
\be
\dis{\int_{R-\varepsilon}^{R+\varepsilon}
dr\delta(r-R)\psi(r)=\frac{1}{2}[\psi(R_+)+\psi(R_-)]}\ \ ,\ \ 
\varepsilon\rightarrow 0\ ,
\ee
we obtain
\be\lb{BC}
\left(\frac{1}{2}G + \hbar c\tau_o\right)\psi(R_+)+
\left(\frac{1}{2}G - \hbar c\tau_o\right)\psi(R_-) =0\ \ ,\ \ 
\tau_o = \left(
\ba{cc} 
0 & -1\cr
1& 0
\ea
\right)\ .
\ee

\vspace{5pt}

\noindent
{\bf Proposition\/}: The general boundary conditions Eq.~(\ref{BC})
define a self-adjoint extension of $\dot{h_{jl}}$ iff $G^+=G$.

\noindent
{\bf Proof}. The proof proceeds in the same manner as that of Proposition 4.1 
in Ref.~\refcite{desm}.

\medskip

To solve the problem, henceforth we shall work in a given total angular
momentum sector $(j\ell)$ of the Dirac spectrum, since the considered
scalar potential is spherically symmetric. Hence, we shall use the
standard notation as follows,
\begin{equation}
\ell=j\pm\frac{1}{2}\ \ \ ,\ \ \
\kappa=\kappa_{jl}=(-1)^{j-\ell+1/2}(j+\frac{1}{2})\ \ \ ,\ \ \
j=1/2,3/2,5/2,\cdots\ .
\end{equation}
In what follows, we shall give expressions valid whether for
$\ell=j+1/2$ or for
$\ell=j-1/2$, the upper/lower signs corresponding to these two situations
in the same order, namely $\ell=j\pm 1/2$. Thus we have
\begin{equation}
\begin{array}{r c l}
\ell=j+\frac{1}{2}&:&\ \ \ \kappa=j+\frac{1}{2}=\ell\ ,\\
& & \\
\ell=j-\frac{1}{2}&:&\ \ \ \kappa=-(j+\frac{1}{2})=-(\ell+1)\ .
\end{array}
\end{equation} 

The strategy for relating appropriate choices of boundary conditions
which define self-adjoint extensions of the Dirac Hamiltonian to the
parameters $\mu_{ij}(z)$ in Krein's formula (see Eq.~(\ref{kf})) in
the general setting is clear. 
It suffices to apply this formula to a set of functions $(F(r')\ G(r')$ 
and determine which boundary conditions are obeyed, 
as a function of the choice for the quantities $\mu_{ij}(z)$. Here, 
we shall not pursue the general analysis, but only restrict to 
the two cases which are called\cite{t} ``boundary conditions of 
the first type", namely when either one of the two functions $f(r)$ or $g(r)$ 
is continuous at $r=R$ while the other is discontinuous at $r=R$ 
with a discontinuity that is proportional to the value of 
the continuous function at that point. Namely, we shall 
restrict to the following two cases,
\begin{equation}
f(R_+)=f(R_-)=f(R)\ \ \ ,\ \ \
g(R_+)-g(R_-)=\alpha f(R)\ ,
\label{eq:bc1}
\end{equation}
\begin{equation}
f(R_+)-f(R_-)=\beta g(R)\ \ \ ,\ \ \
g(R_+)=g(R_-)=g(R)\ ,
\label{eq:bc2}
\end{equation}
where $\alpha={\alpha_o}/(\hbar c)$ and 
$\beta=-{\beta_o}/(\hbar c)$ are arbitrary real parameters.
As we shall see, since such a choice of boundary conditions may indeed
be put into one-to-one correspondence with a choice for the Krein parameters
$\mu_{ij}(z)$, these conditions define two separate one-parameter
self-adjoint extensions of the Dirac Hamiltonian, associated to the
$\delta$-sphere scalar potential.

{}From now on, the boundary conditions defined by Eq.~(\ref{eq:bc1})
will be called ``$\alpha$-type boundary condition of the first
type'' and those defined by Eq.~(\ref{eq:bc2}) will be
called ``$\beta$-type boundary condition of the first type''.
Note, as it should be, that these boundary conditions are particular 
cases of boundary conditions given in Eq.~(\ref{BC}).

The self-adjoint extension of $\dot{H}$ corresponding to the
specific boundary conditions in Eqs.~(\ref{eq:bc1}) and (\ref{eq:bc2})
write, respectively,
\be 
H_{\alpha} = \bigoplus_{j=\frac{1}{2}}^{\infty}\;
\bigoplus_{l=j-\frac{1}{2}}^{j+\frac{1}{2}}\;
U_{j\ell}^{-1} h_{{j\ell},\alpha}U_{j\ell}\otimes {\U }\ \ ,\ \ 
\alpha = \{{\alpha}_{j\ell}\}\ ,
\ee
with
\bea \nonumber
{h}_{j\ell,{\alpha}_\ell} &\equiv & \tau\ ,\\ 
{\mathcal D}(h_{{j\ell},\alpha_\ell}) &=& \left\{\psi\in L^2((0,\infty))\otimes 
\C^2\;;\; f_{j\ell}\in AC_{loc}((0,\infty))\right. ;\nonumber\\ 
& & g_{j\ell}\in AC_{loc}((0,\infty) \setminus \{R\}) ;\nonumber\\ 
& & g_{j\ell}(R_+) - g_{j\ell}(R_-) = \alpha f_{j\ell}(R) ;\nonumber\\ 
& & \left. h_{{j\ell},\alpha}\psi \in L^2((0,\infty))
\otimes \C^2\; \right\}\ , 
\eea
and
\be 
H_{\beta} = \bigoplus_{j=\frac{1}{2}}^{\infty}\;
\bigoplus_{l=j-\frac{1}{2}}^ 
{j+\frac{1}{2}}\; U_{j\ell}^{-1} h_{{j\ell},\beta}U_{j\ell}
\otimes {\U }\ \ ,\ \ 
\beta = \{{\beta}_{j\ell}\}\ , 
\ee
with
\bea \nonumber
{h}_{j\ell,{\alpha}_\ell} &\equiv & \tau\ ,\\  
{\mathcal D}(h_{{j\ell},\beta_{j\ell}}) &=& \left\{\psi\in L^2((0,\infty))\otimes 
\C^2\;;\; g_{j\ell}\in AC_{loc}((0,\infty))\right. ;\nonumber\\ 
& & f_{j\ell}\in AC_{loc}((0,\infty) \setminus \{R\}) ;\nonumber\\ 
& & f_{j\ell}(z;R_+) - f_{j\ell}(z;R_-) = \beta_{j\ell}g_{j\ell}(z;R) 
 ;\nonumber\\ 
& & \left. h_{{j\ell},\beta_{j\ell}}\psi \in L^2((0,\infty))
\otimes \C^2\;\right\}\ . 
\eea

\section{Solutions to the Free Radial Dirac Hamiltonian}

In the present section, we are interested in the determination
of the eigenstates of the free radial operator, and in particular their
absolute normalisation as required for the construction of the
associated free Dirac resolvent or kernel of the corresponding 
Green's function. Hence, we must solve for the eigenvalue equation
\begin{equation}
\left(\begin{array}{c c}
mc^2 & \hbar c\left(-\frac{d}{dr}+\frac{\kappa}{r}\right) \\
\hbar c\left(\frac{d}{dr}+\frac{\kappa}{r}\right) & -mc^2
\end{array}\right)
\left(\begin{array}{c}
f(r) \\ g(r) \end{array}\right)=
E\left(\begin{array}{c}
f(r) \\ g(r) \end{array}\right)\ .
\end{equation}
The choices of normalisation and signs for these components is standard
and we required only that
\begin{equation}
\int_0^\infty dr\left(f^*(r;p)\ \ g^*(r;p)\right)
\left(\begin{array}{c}
f(r;p') \\ g(r;p')\end{array}\right)=\delta(p-p')\ .
\end{equation}
We have to distinguish the two cases of either
positive or negative energy states, since
we know that the free Dirac energy spectrum is that of
positive and negative energy states such that
\begin{equation}
E=\pm\sqrt{(pc)^2+(mc^2)^2}\ \ \ ,\ \ \ p\ge 0\ .
\end{equation}
Then, introducing the following parametrisations,
\begin{equation}
E(p)=\pm\omega(p)\ \ \ ,\ \ \
\omega(p)=\sqrt{(pc)^2+(mc^2)^2}\ \ \ ,\ \ \
\rho=\frac{pc}{\hbar c}r\ ,
\end{equation}
and labelling all distinct eigenstates by the value for $p>0$ and the
sign of the energy, one finds for the normalised solutions to the free radial
Dirac Hamiltonian of positive energy, for $\ell=j\pm 1/2$,
\begin{equation}
f_+(r;p)=\frac{pc}{\sqrt{2\hbar\omega(p)(\omega(p)-mc^2)}}
\left(\frac{pr}{\hbar}\right)^{1/2}J_{\ell+1/2}\left(\frac{pr}{\hbar}\right)\ ,
\end{equation}
\begin{equation}
g_+(r;p)=\frac{pc}{\sqrt{2\hbar\omega(p)(\omega(p)+mc^2)}}
\left(\frac{pr}{\hbar}\right)^{1/2}J_{\ell+1/2\mp 1}
\left(\frac{pr}{\hbar}\right)\ .
\end{equation}
The negative energy solutions are given by, with $\ell=j\pm 1/2$,
\begin{equation}
f_-(r;p)=\frac{pc}{\sqrt{2\hbar\omega(p)(\omega(p)+mc^2)}}
\left(\frac{pr}{\hbar}\right)^{1/2}J_{\ell+1/2}\left(\frac{pr}{\hbar}\right)\ ,
\end{equation}
\begin{equation}
g_-(r;p)=\frac{-pc}{\sqrt{2\hbar\omega(p)(\omega(p)-mc^2)}}
\left(\frac{pr}{\hbar}\right)^{1/2}J_{\ell+1/2\mp 1}
\left(\frac{pr}{\hbar}\right)\ .
\end{equation}
Here, $J_{\mu}(\cdot)$ are Bessel functions of order $\mu$.\cite{as}
Finally, let us note that one may readily check that the
configurations of opposite energy have a vanishing inner product, given
the negative sign in the function $g_-(r;p)$ for the negative energy
solutions relative to that of the positive energy ones. Hence, we have
constructed a complete basis of eigenstates for the free radial Dirac
Hamiltonian, in terms of which it is thus possible to construct the
explicit expressions for the associated resolvent.

\section{The Resolvent of the Free Dirac Radial Hamiltonian}

In an abstract and formal manner, let us consider a self-adjoint
operator $A$ with spectrum\footnote{We assume here that the spectrum
is not degenerate, since this is the situation encountered presently. The case
of a degenerate spectrum is a straightforward generalisation.}
\begin{equation}
A|\psi_n>=\lambda_n|\psi_n>\ ,
\end{equation}
where the eigenstates are orthonormalised,
\begin{equation}
<\psi_n|\psi_m>=\delta_{n,m}\ .
\end{equation}
Consequently, given a complex parameter $z$, the associated resolvent
$\left[A-z\right]^{-1}$ is simply given by the representation
\begin{equation}
\left[A-z\right]^{-1}=\sum_n|\psi_n>\frac{1}{\lambda_n-z}<\psi_n|\ ,
\end{equation}
where $z$ does not belong to the $A$-spectrum. We shall simply follow
exactly the same construction for the $(r,r')$ kernel of the free Dirac
radial Hamiltonian, since we have constructed its spectrum in the previous
section. In our case, this resolvent is also a 2$\times$2 matrix, each
of which elements are obtained in a likewise manner.
Hence, we have for the resolvent kernel of the free Dirac radial Hamiltonian
\begin{equation}
G_0(r,r';z)=\left(\begin{array}{c c}
G_{11}(r,r';z) & G_{12}(r,r';z) \\
G_{21}(r,r';z) & G_{22}(r,r';z) \end{array}\right)\ .
\end{equation}

Introducing the quantity $k=k(z)$ defined by
\begin{equation}
k(z)c=\sqrt{z^2-(mc^2)^2}\ ,
\end{equation}
with the usual branch cut along the negative real axis for the square-root
function, we have for the $G_{11}$ element
\begin{equation}\label{e1}
 G_{11}(r,r';z)=
\frac{i\pi}{2}\frac{z+mc^2}{(\hbar c)^2}\sqrt{rr'}
H^{(1)}_{\ell+1/2}\left(\frac{k(z)c}{\hbar c}r\right)\
J_{\ell+1/2}\left(\frac{k(z)c}{\hbar c}r'\right)\ ,
\end{equation}
\begin{equation}\label{e2}
 G_{11}(r,r';z)=
\frac{i\pi}{2}\frac{z+mc^2}{(\hbar c)^2}\sqrt{rr'}
J_{\ell+1/2}\left(\frac{k(z)c}{\hbar c}r\right)\
H^{(1)}_{\ell+1/2}\left(\frac{k(z)c}{\hbar c}r'\right)\ ,
\end{equation}
Eqs.~(\ref{e1}) and (\ref{e2}) being valid for $r>r'$ and $r<r'$,
respectively. Note that as a consequence, the quantity $G_{11}(r,r';z)$ 
is also well defined for $r=r'$, as it should.

The second diagonal element is given by
\be
\label{e3}
G_{22}(r,r';z)=
\frac{i\pi}{2}\frac{z-mc^2}{(\hbar c)^2}\sqrt{rr'}
H^{(1)}_{\ell+1/2\mp 1}\left(\frac{k(z)c}{\hbar c}r\right)\
J_{\ell+1/2\mp 1}\left(\frac{k(z)c}{\hbar c}r'\right)\ ,
\ee
\begin{equation}
\label{e4}
G_{22}(r,r';z)=
\frac{i\pi}{2}\frac{z-mc^2}{(\hbar c)^2}\sqrt{rr'}
J_{\ell+1/2\mp 1}\left(\frac{k(z)c}{\hbar c}r\right)\
H^{(1)}_{\ell+1/2\mp 1}\left(\frac{k(z)c}{\hbar c}r'\right)\ ,
\end{equation}
Eqs.~(\ref{e3}) and (\ref{e4}) being valid for $r>r'$ and $r<r'$,
respectively.

For the off-diagonal element $G_{12}(r,r';z)$, we obtain
the following two expressions valid for $r>r'$ and $r<r'$,
respectively,
\begin{equation}
G_{12}(r,r';z)=
\frac{i\pi}{2}\frac{k(z)c}{(\hbar c)^2}\sqrt{rr'}
H^{(1)}_{\ell+1/2}\left(\frac{k(z)c}{\hbar c}r\right)\
J_{\ell+1/2\mp 1}\left(\frac{k(z)c}{\hbar c}r'\right)\ ,
\end{equation}
\begin{equation}
G_{12}(r,r';z)=
\frac{i\pi}{2}\frac{k(z)c}{(\hbar c)^2}\sqrt{rr'}
J_{\ell+1/2}\left(\frac{k(z)c}{\hbar c}r\right)\
H^{(1)}_{\ell+1/2\mp 1}\left(\frac{k(z)c}{\hbar c}r'\right)\ .
\end{equation}
Finally for $G_{21}(r,r';z)$ element, we obtain
\begin{equation}\label{e5}
G_{21}(r,r';z)=
\frac{i\pi}{2}\frac{k(z)c}{(\hbar c)^2}\sqrt{rr'}
H^{(1)}_{\ell+1/2\mp 1}\left(\frac{k(z)c}{\hbar c}r\right)\
J_{\ell+1/2}\left(\frac{k(z)c}{\hbar c}r'\right)\ ,
\end{equation}
\begin{equation}\label{e6}
G_{21}(r,r';z)=
\frac{i\pi}{2}\frac{k(z)c}{(\hbar c)^2}\sqrt{rr'}
J_{\ell+1/2\mp 1}\left(\frac{k(z)c}{\hbar c}r\right)\
H^{(1)}_{\ell+1/2}\left(\frac{k(z)c}{\hbar c}r'\right)\ ,
\end{equation}
Eqs.~(\ref{e5}) and (\ref{e6}) being valid for $r>r'$ and $r<r'$,
respectively.

\section{The Deficient Index Subspaces}

In order to use Krein's formula to express the Dirac radial resolvent
for a given choice of boundary conditions which define a self-adjoint
extension of the Dirac Hamiltonian related to a $\delta$-sphere scalar
potential, it is first necessary to establish a basis for the deficiency
subspaces. This amounts to identifying the eigensolutions to the previous
eigenvalue problem in which the energy eigenvalue $E$ is now replaced
by a complex parameter $z$ which does not belong to the spectrum of
the free and interacting Hamiltonians. The only restriction is that the
obtained solutions must be normalisable for the norm $\int_0^\infty dr$.

Consequently, for a given complex value of $z$, let us introduce
again the quantity
\begin{equation}
k(z)c=\sqrt{z^2-(mc^2)^2}\ ,
\end{equation}
where the square-root function is defined as previously.
It is then straightforward to show that there exist two unique
linearly independent normalisable solutions, for Im$\,k(z)>0$.
The first solution is given by,
\begin{equation}
\begin{array}{r l}
{\rm .\ if\ }r<R:&
f_1(r;z)=\frac{k(z)c}{\sqrt{2\hbar z(z-mc^2)}}
\left(\frac{k(z)c}{\hbar c}r\right)^{1/2}J_{\ell+1/2}
\left(\frac{k(z)c}{\hbar c}r\right)\ ,\\
& g_1(r;z)=\frac{k(z)c}{\sqrt{2\hbar z(z+mc^2)}}
\left(\frac{k(z)c}{\hbar c}r\right)^{1/2}J_{\ell+1/2\mp 1}
\left(\frac{k(z)c}{\hbar c}r\right)\ ;\\
 & \\
{\rm .\ if\ }r>R:&
f_1(r;z)=0\ ,\\
& g_1(r;z)=0\ .
\end{array}
\end{equation}
The second solution is given by
\begin{equation}
\begin{array}{r l}
{\rm .\ if\ }r<R:&
f_2(r;z)=0\ ,\\
& g_2(r;z)=0\ ;\\
 & \\
{\rm .\ if\ }r>R:&
f_2(r;z)=\frac{k(z)c}{\sqrt{2\hbar z(z-mc^2)}}
\left(\frac{k(z)c}{\hbar c}r\right)^{1/2}H^{(1)}_{\ell+1/2}
\left(\frac{k(z)c}{\hbar c}r\right)\ ,\\
& g_2(r;z)=\frac{k(z)c}{\sqrt{2\hbar z(z+mc^2)}}
\left(\frac{k(z)c}{\hbar c}r\right)^{1/2}H^{(1)}_{\ell+1/2\mp 1}
\left(\frac{k(z)c}{\hbar c}r\right)\ .
\end{array}
\end{equation}

If Im$\,k(z)<0$, the corresponding deficiency subspace is spanned by
the complex conjugates of the above expressions, except for the fact
that $k(z)$ is then still given by the above definition in terms of $z$
rather than in terms of $z^*=\bar{z}$. In other words, for Im$\,k(z)<0$,
the deficiency subspace is spanned by the same solutions $(f_1(r;z)\ g_1(r;z))$
as above and by functions $(f'_2(r;z)\ g'_2(r;z))$ which are given by the
same expressions as those for $(f_2(r;z)\ g_2(r;z))$ with the
function $H^{(1)}_{\ell+1/2\mp 1}$ replaced by $H^{(2)}_{\ell+1/2\mp 1}$.

Since Krein's formula reads\cite{ag}
\begin{equation}\lb{kf}
G(r,r';z)=G_0(r,r';z)+\sum_{i,j=1}^2\mu_{ij}(z)
\left(\begin{array}{c}
f_i(r;z) \\ g_i(r;z)\end{array}\right)
\left(f^*_j(r';\bar{z})\ \ g^*_j(r';\bar{z})\right)\ ,
\end{equation}
where $G(r,r';z)$ stands for the resolvent kernel of the interacting
Hamiltonian, it is clear that the specific values for the coefficients
$\mu_{ij}(z)$ are defined up to a normalisation which is correlated to
the chosen normalisation for the solutions $(f_i\ g_i)$ constructed
above. As mentioned already, the latter normalisation is simply chosen
by analogy with that of the free energy eigenstates, and it might thus
turn out that there could be a choice of normalisation rendering
the expressions for the coefficients $\mu_{ij}(z)$ simpler.
This is indeed the case, but in a manner which is dependent on the
specific choice of boundary conditions which is associated to a specific
self-adjoint extension of the Dirac Hamiltonian. It is for this reason
that we have kept the above normalisation, leaving it as an exercise
to change the normalisation to make it simpler once a specific choice
of self-adjoint boundary conditions is considered. This is rather
straightforward on the basis of the results to be presented in the next
section.

Note also that since the deficiency indices of the Hamiltonian are $(2,2)$,
the set of its self-adjoint extensions is characterized by four independent
real parameters, which are thus in correspondence with the four parameters
$\mu_{ij}(z)$. In other words, any choice of boundary conditions for the
functions $f(r)$ and $g(r)$ obeying the interacting Hamiltonian energy
eigenvalue problem which may be put into unique correspondence with the
parameters $\mu_{ij}(z)$ in Krein's formula determines a self-adjoint
extension of the Dirac Hamiltonian. The next section displays two such
choices, each characterized by a single real parameter, leaving it
aside how to construct the general case and how to identify its relationship
to $\delta$-sphere scalar and vector interactions.

\section{The Resolvent Equation}

\subsection{The $\alpha$-Type Boundary Condition of the First Type}

Requiring $f(r)$ to be continuous at $r=R$ implies that we must have
\begin{equation}
\mu_{11}(z)=\mu_1(z) H^{(1)}_{\ell+1/2}(\frac{k(z)R}{\hbar})\ \ ,\ \
\mu_{21}(z)=\mu_1(z) J_{\ell+1/2}(\frac{k(z)R}{\hbar})\ ,
\end{equation}
\begin{equation}
\mu_{12}(z)=\mu_2(z) H^{(1)}_{\ell+1/2}(\frac{k(z)R}{\hbar})\ \ ,\ \
\mu_{22}(z)=\mu_2(z) J_{\ell+1/2}(\frac{k(z)R}{\hbar})\ ,
\end{equation}
where $\mu_1(z)$ and $\mu_2(z)$ are real parameters still to be determined.
After some algebra, one finds that the boundary condition involving
the difference $g(R_+)-g(R_-)$ implies the following relations for the
quantities $\mu_1(z)$ and $\mu_2(z)$ in terms of $\alpha$, for $\ell=j\pm 1/2$,
\bea
\mu_1(z)=&&\frac{i\pi}{2}\frac{R(z+mc^2)}{\hbar c}
\frac{i\pi}{2}\frac{2\hbar z}{\hbar c}\frac{1}{k(z)c}
\times\nonumber\\
&\times&\frac{\alpha H^{(1)}_{\ell+1/2}(\frac{k(z)R}{\hbar})}
{\mp 1-\alpha\frac{i\pi}{2}\frac{R(z+mc^2)}{\hbar c}
H^{(1)}_{\ell+1/2}(\frac{k(z)R}{\hbar})J_{\ell+1/2}(\frac{k(z)R}{\hbar})}\ ,
\eea
\bea
\mu_2(z)=&&\frac{i\pi}{2}\frac{R(z+mc^2)}{\hbar c}
\frac{i\pi}{2}\frac{2\hbar z}{\hbar c}\frac{1}{k(z)c}
\times\nonumber\\
&\times&
\frac{\alpha J_{\ell+1/2}(\frac{k(z)R}{\hbar})}
{\mp 1-\alpha\frac{i\pi}{2}\frac{R(z+mc^2)}{\hbar c}
H^{(1)}_{\ell+1/2}(\frac{k(z)R}{\hbar})J_{\ell+1/2}(\frac{k(z)R}{\hbar})}\ .
\eea

Note that the quantity that multiplies $\alpha$ in the denominator of these
two expressions is simply the value for $G_{11}(R,R;z)$ for the free Dirac
resolvent, in exactly the same manner as for the nonrelativistic
Schr\"odinger problem. Furthermore, it should be clear that most of the
overall normalisation factors multiplying the last fraction could be
absorbed directly through a renormalisation of the functions
$f_i(r;z)$ and $g_i(r;z)$ spanning the deficiency subspace. We shall
leave this point open, since this is rather trivial, even though it simplifies
somewhat the expression for the coefficients $\mu_1(z)$ and $\mu_2(z)$.
Finally, notice that the final expressions for all four coefficients
$\mu_{ij}(z)$ generalise similar expressions relevant to
the nonrelativistic Schr\"odinger problem. In the latter case, all the factors
involving the Bessel functions evaluated at $k(z)R/\hbar$ may be absorbed into
the normalisation of the vectors spanning the deficiency subspace, leaving
over a rather simple expression for the coefficient $\mu(z)$ in that case.
For the Dirac Hamiltonian, this is no longer the case in a simple way,
but still the general structure is maintained for the solutions of the
coefficients $\mu_{ij}(z)$.

Finally the resolvent of the extended Dirac Hamiltonian
for the $\alpha$-type boundary conditions is given by
\be
\left(h_{l,\alpha_l}-z\right)^{-1}=\left(h_{l,0}-z\right)^{-1}
+A(z)\U_2 \psi(r;z)\left(\psi^* (r;\overline{z}),\cdot \right)\ ,
\ee
with the following notations,
\bea
A(z)=&&\frac{i\pi}{2}\frac{R(z+mc^2)}{\hbar c}
\frac{i\pi}{2}\frac{2\hbar z}{\hbar c}\frac{1}{k(z)c}
\times\nonumber\\
&\times&\frac{\alpha }
{\mp 1-\alpha\frac{i\pi}{2}\frac{R(z+mc^2)}{\hbar c}
H^{(1)}_{\ell+1/2}(\frac{k(z)R}{\hbar})
J_{\ell+1/2}(\frac{k(z)R}{\hbar})}\ ,
\eea
\be
\psi(r;z)=\left(
\ba{l}
\psi_1(r;z)
\\
\psi_2(r;z)
\ea
 \right)\ ,
\ee
\bea
\psi_1(r;z)&=&
a(z)\times \left(\frac{k(z)}{\hbar}r\right)^{1/2}
\left[J_{\ell+1/2}(\frac{k(z)r}
{\hbar})H^{(1)}_{\ell+1/2}(\frac{k(z)R}{\hbar})+
\right.\nonumber\\
&&+ \left.
J_{\ell+1/2}(\frac{k(z)R}{\hbar})H^{(1)}_{\ell+1/2}
(\frac{k(z)r}{\hbar}) \right]\ ,
\eea
\bea
\psi_2(r;z)&=&
b(z)\times \left(\frac{k(z)}{\hbar}r\right)^{1/2}
\left[J_{\ell+1/2\mp 1}(\frac{k(z)r}
{\hbar})H^{(1)}_{\ell+1/2}(\frac{k(z)R}{\hbar})+
\right.\nonumber\\
&&+\left.
 J_{\ell+1/2}(\frac{k(z)R}{\hbar})H^{(1)}_{\ell+1/2\mp 1}
(\frac{k(z)r}{\hbar}) \right]\ ,
\eea
\be\lb{parame}
a(z)=\frac{k(z)c}{\sqrt{2\hbar z(z-mc^2)}}\ \ ,\ \ 
b(z)=\frac{k(z)c}{\sqrt{2\hbar z(z+mc^2)}}\ .
\ee

\subsection{The $\beta$-Type Boundary Condition of the First Type}
\label{Subsect6.2}

A similar analysis is possible when the continuity requirement is
imposed for $g(r)$,
\begin{equation}
g(R_+)=g(R_-)=g(R)\ .
\end{equation}
It then turns out that the $\beta$-type boundary conditions of the
first type, Eq.~(\ref{eq:bc2}), are then possible, and provide thus 
a one-parameter self-adjoint extension of the Dirac Hamiltonian. 
In this case, we must have
\begin{equation}
\mu_{11}(z)=\mu_1(z) H^{(1)}_{\ell+1/2\mp 1}(\frac{k(z)R}{\hbar})\ \ ,\ \
\mu_{21}(z)=\mu_1(z) J_{\ell+1/2\mp 1}(\frac{k(z)R}{\hbar})\ ,
\end{equation}
\begin{equation}
\mu_{12}(z)=\mu_2(z) H^{(1)}_{\ell+1/2\mp 1}(\frac{k(z)R}{\hbar})\ \ ,\ \
\mu_{22}(z)=\mu_2(z) J_{\ell+1/2\mp 1}(\frac{k(z)R}{\hbar})\ ,
\end{equation}
while the coefficients $\mu_1(z)$ and $\mu_2(z)$ are given as follows
in terms of the real parameter $\beta$ defining the boundary conditions
Eq.~(\ref{eq:bc2}), for $\ell=j\pm 1/2$,
\bea
\mu_1(z)=&&\frac{i\pi}{2}\frac{R(z-mc^2)}{\hbar c}
\frac{i\pi}{2}\frac{2\hbar z}{\hbar c}\frac{1}{k(z)c}
\times\nonumber\\
&\times&\frac{\beta H^{(1)}_{\ell+1/2\mp 1}(\frac{k(z)R}{\hbar})}
{\pm 1-\beta\frac{i\pi}{2}\frac{R(z-mc^2)}{\hbar c}
H^{(1)}_{\ell+1/2\mp 1}(\frac{k(z)R}{\hbar})
J_{\ell+1/2\mp 1}(\frac{k(z)R}{\hbar})}\ ,
\eea
\bea
\mu_2(z)=&&\frac{i\pi}{2}\frac{R(z-mc^2)}{\hbar c}
\frac{i\pi}{2}\frac{2\hbar z}{\hbar c}\frac{1}{k(z)c}
\times\nonumber\\
&\times&\frac{\beta J_{\ell+1/2\mp 1}(\frac{k(z)R}{\hbar})}
{\pm 1-\beta\frac{i\pi}{2}\frac{R(z-mc^2)}{\hbar c}
H^{(1)}_{\ell+1/2\mp 1}(\frac{k(z)R}{\hbar})
J_{\ell+1/2\mp 1}(\frac{k(z)R}{\hbar})}\ .
\eea

The same remarks as those made for the $\alpha$-type boundary conditions are
of application in this case, of course, with in particular the value for
$G_{22}(R,R;z)$ explicitly multiplying the parameter $\beta$ in the
denominator of the last fraction factor. Finally, the resolvent in
a compact form, for the $\beta$-type boundary conditions,
is given by
\be
\left(h_{l,\beta_l}-z\right)^{-1}=\left(h_{l,0}-z\right)^{-1}
+\widetilde{A}(z)\U_2 \widetilde{\psi}(r;z)
\left(\widetilde{\psi}^* (r;\overline{z}),\cdot \right)\ ,
\ee
with the notations
\bea
 \widetilde{\psi}(r;z)
=\left(
\ba{l}
\widetilde{\psi}_1(r;z)
\\
\widetilde{\psi}_2(r;z)
\ea
 \right)\ ,
\eea
\bea
\widetilde{\psi}_1(r;z)&=&
a(z)\times \left(\frac{k(z)}{\hbar}r\right)^{1/2}
\left[J_{\ell+1/2}(\frac{k(z)r}{\hbar})
H^{(1)}_{\ell+1/2\mp 1}(\frac{k(z)R}{\hbar})+
\right.\nonumber\\
&&+\left.
 J_{\ell+1/2\mp 1}(\frac{k(z)R}{\hbar})
H^{(1)}_{\ell+1/2}(\frac{k(z)r}{\hbar}) \right]\ ,
\eea
\bea
\widetilde{\psi}_2(r;z)&=&
b(z)\times \left(\frac{k(z)}{\hbar}r\right)^{1/2}
\left[J_{\ell+1/2\mp 1}(\frac{k(z)r}{\hbar})
H^{(1)}_{\ell+1/2\mp 1}(\frac{k(z)R}{\hbar})+
\right.\nonumber\\
&&+\left.
 J_{\ell+1/2\mp 1}(\frac{k(z)R}{\hbar})
H^{(1)}_{\ell+1/2\mp 1}(\frac{k(z)r}{\hbar}) \right]\ ,
\eea
\bea
\widetilde{A}(z)=&&\frac{i\pi}{2}\frac{R(z-mc^2)}{\hbar c}
\frac{i\pi}{2}\frac{2\hbar z}{\hbar c}\frac{1}{k(z)c}
\times\nonumber\\
&\times&\frac{\beta }
{\pm 1-\beta\frac{i\pi}{2}\frac{R(z-mc^2)}{\hbar c}
H^{(1)}_{\ell+1/2\mp 1}(\frac{k(z)R}{\hbar})J_{\ell+1/2\mp 1}
(\frac{k(z)R}{\hbar})}\ .
\eea

\section{Conclusion}

We have thus constructed the resolvent kernel $G(r,r';z)$
for the one-parameter self-adjoint extensions of the Dirac Hamiltonian
associated to the $\delta$-sphere scalar potential and characterized
by the $\alpha$- and $\beta$-type boundary conditions of the first type
defined in Eqs.~(\ref{eq:bc1}) and (\ref{eq:bc2}). More general cases could
be considered on the basis of the general expressions obtained above
for the values of $f(R_\pm)$ and $g(R_\pm)$, in order to identify all
possible self-adjoint extensions parametrized by four independent real
parameters.

Associated to the boundary conditions Eqs.~(\ref{eq:bc1}) and (\ref{eq:bc2}), 
these results should now enable the analysis of the scattering, spectral
and resonance properties of the constructed self-adjoint
extensions of the Dirac operator.\cite{t}

Finally, let us say that it is also possible to construct a subclass of
self-adjoint extensions characterized by two parameters.
Shabani and Vyabandi attempted such a construction
in their work\cite{sv} but it is not clear to us how
the resolvent kernel of the free Dirac operator which they obtained 
and on which the remainder of their analysis rests in a crucial way,
compares to its explicit contruction as detailed in the present
contribution.

\section*{Acknowledgements}

The authors acknowledge the Belgian Cooperation CUD-CIUF/UAC
for financial support. The work of JG is partially supported by the 
Federal Office for Scientific, Technical and Cultural Affairs (Belgium) 
through the Interuniversity Attraction Pole P5/27.

\end{document}